\documentclass[aps,prl,reprint,superscriptaddress,nobibnotes,amsmath,amssymb]{revtex4-2}

\usepackage[utf8]{inputenc}
\usepackage{mathtools,bm}
\usepackage{graphicx}
\usepackage{xcolor}
\usepackage{tikz}
\usetikzlibrary{arrows.meta,calc,decorations.pathreplacing}
\usepackage{hyperref}
\definecolor{defect}{HTML}{176EAA}
\definecolor{bankupper}{HTML}{B83246}
\definecolor{banklower}{HTML}{C47A13}
\definecolor{ink}{HTML}{253241}
\newcommand{\dd}{\mathrm d}
\newcommand{\Tr}{\operatorname{Tr}}
\newcommand{\clr}{c_{LR}}

\newcommand{\BKM}{\mathrm{BKM}}
\newcommand{\SM}{Supplemental Material~\cite{SM}}

\begin{document}

\title{Entanglement Across A Conformal Interface Is Bounded Below By Energy Transmission}

\author{Elisa Gramaglia}
\email{elisa.gramaglia@sissa.it}
\affiliation{SISSA, Via Bonomea 265, 34136 Trieste, Italy}
\affiliation{INFN, Sezione di Trieste, Via Valerio 2, 34127 Trieste, Italy}

\author{Andreas Karch}
\email{karcha@utexas.edu}
\affiliation{Theory Group, Weinberg Institute, Department of Physics, University of Texas, 2515 Speedway, Austin, TX 78712, USA}

\author{Hirosi Ooguri}
\email{ooguri@caltech.edu}
\affiliation{Walter Burke Institute for Theoretical Physics \& Leinweber Forum for Theoretical Physics \\
California Institute of Technology, Pasadena, CA 91125, USA}
\affiliation{Kavli Institute for the Physics and Mathematics of the Universe, WPI \\ University of Tokyo, Kashiwa 277-8583, Japan}

\author{Mianqi Wang}
\email{mianqi.wang@ipmu.jp}
\affiliation{Kavli Institute for the Physics and Mathematics of the Universe, WPI \\ University of Tokyo, Kashiwa 277-8583, Japan}

\date{\today}

\begin{abstract}
We prove the inequality $c_{LR}\leq c_{\mathrm{eff}}$ for generic conformal interfaces in any compact unitary conformal field theory in two dimensions. It relates the energy transmission coefficient $c_{LR}$ and the effective central charge $c_{\mathrm{eff}}$, which measures the entanglement across an interface. With a known upper bound for $c_{\mathrm{eff}}$, it gives $c_{LR}\leq c_{\mathrm{eff}}\leq\min(c_L,c_R)$, where $c_L$ and $c_R$ are the central charges on either side. The key step is relating $c_{LR}$ to the mixed second derivative of the entanglement entropy with respect to the endpoints of intervals symmetric about the interface, and applying strong subadditivity.
\end{abstract}

\maketitle

\section{Introduction}

The purpose of this Letter is to compare entanglement and energy transmission across conformal interfaces in two-dimensional unitary conformal field theories. The effective central charge $c_{\mathrm{eff}}$  governs the logarithmic contribution to entanglement entropy \cite{SakaiSatoh2008Entanglement,BrehmBrunner2015Ising}, while the coefficient $c_{LR}$   of the stress-tensor two-point function determines energy transmission \cite{quella2007reflection}. For an interface joining theories with central charges $c_L$ and $c_R$, our previous paper \cite{karch2023universalityeffectivecentralcharge} with Kusuki and Sun derived the upper bound $c_{\mathrm{eff}}\leq\min(c_L,c_R)$. 
In the other paper \cite{karch2024universalbound} with Kusuki and Sun, we 
conjectured the complementary lower bound $c_{LR}\leq c_{\mathrm{eff}}$  and presented evidence from holographic examples, conformal perturbation theory, and free field theories.

In this Letter, we prove the conjectured lower bound $c_{LR}\leq c_{\mathrm{eff}}$. Together with the upper bound of \cite{karch2023universalityeffectivecentralcharge}, this establishes
\[
c_{LR}\leq c_{\mathrm{eff}}\leq\min(c_L,c_R).
\]
The key step is to relate $c_{LR}$  to the second-order variation of the entanglement entropy about an interval centered on the interface. We evaluate this response using the interface displacement operator and then apply strong subadditivity to bound it by $c_{\mathrm{eff}}$.

Consider a conformal interface at $x=0$ between two unitary conformal field theories (CFTs) with central charges $c_L$ and $c_R$\footnote{We consider a conformal interface between two unitary CFTs and assume that the full theory, including the interface, is reflection positive. We work in a fixed vacuum sector and use regulators preserving this positivity.}. Normalize the transmission
coefficient  \cite{Meineri:2019ycm} by
\begin{align}
  \langle T_L(z)T_R(z')\rangle_{\mathcal I}
  &=\frac{c_{LR}}{2(z-z')^4},\nonumber\\
  \langle T_\alpha(z)T_\alpha(z')\rangle
  &=\frac{c_\alpha}{2(z-z')^4},\quad \alpha=L,R,
  \label{eq:TT-normalization}
\end{align}
for spacelike separated points with $z=x-t$ in Lorentzian signature.
For an interval $[-a,b]$, parametrized as $a=\ell e^{-u}$ and $b=\ell e^u$ and with a common proper
endpoint cutoff $\epsilon$, its entanglement entropy in the vacuum interface state is
\begin{equation}
  S(a,b)=\frac{c_L}{6}\log\frac{2a}{\epsilon}
        +\frac{c_R}{6}\log\frac{2b}{\epsilon}+F(u),
  \label{eq:entropy-u}
\end{equation}
up to corrections vanishing as $\epsilon\to0$ at fixed $a,b$.
After folding $\mathrm{CFT}_R$ onto $x<0$, the entropy is computed
by twist fields $\sigma_L$ at $x=-a$ and $\sigma_R$ at $x=-b$ in a
boundary CFT. The boundary-preserving inversion $z\mapsto\ell^2/z$
exchanges the two positions while retaining the labels, and its
conformal factors $b/a$ and $a/b$ at the insertions rescale the
cutoffs so as to reproduce the endpoint terms in
\eqref{eq:entropy-u}. Hence $F(u)=F(-u)$, {\it i.e.} $F$ depends only on the cross-ratio
$\chi=(a+b)^2/(4ab)=\cosh^2u$.
In terms of $\ell,u$, \eqref{eq:entropy-u} becomes \footnote{With the same endpoint subtraction
and additive normalization, $F(u)=\log g_c^{(2)}(e^{-2u})$ in
\cite{afxonidis2026connectingboundaryentropyeffective,afxonidis2025boundaryentropyfunctioninterface}. At $u=0$ the interval folds onto an interval of length $a$ ending on the boundary, so with this normalization $F(0)=\log g$, the interface entropy. For $c_L=c_R=c$, writing instead $S=\frac{c}{3}\log\frac{a+b}{\epsilon}+\log g(\chi)$ gives $F(u)=\log g(\chi)+\frac{c}{6}\log\chi$.}
\begin{equation}
  S(u)
  =
  \frac{c_L+c_R}{6}\log\frac{2\ell}{\epsilon}
  +
  \frac{c_R-c_L}{6}u
  +
  F(u).
  \label{eq:entropy-expanded-u}
\end{equation}

Below, we will derive the following relation between $c_{LR}$ and the second-order variation of the entanglement entropy at the symmetric interval ($u\to 0$):
\begin{equation}
 F(u)-F(0)=\frac{F''(0)}{2}u^2+o(u^2)=\frac{c_{LR}}{6}u^2+o(u^2).
  \label{eq:curvature-target-expansion}
\end{equation}

On the other hand, $c_\text{eff}$ is obtained in the large-asymmetric limit of the entanglement entropy $S(u)$ at $|u|\to\infty$ \footnote{Note that throughout this paper, we first take the cutoff $\epsilon\to 0$ at fixed $u$, then take the large asymmetric limit $u\to\infty$ while keeping $\epsilon\ll a$.}, where we have
  $F(u)=\frac{1}{3}c_{\mathrm{eff}}|u|+o(|u|)$ \cite{karch2023universalityeffectivecentralcharge,afxonidis2026connectingboundaryentropyeffective,wang2026twistoperatorbopeentanglement}.

Using the above identities and strong subadditivity, we can prove the conjectured bound $c_{LR}\le c_{\text{eff}}$ for any conformal interface in any compact unitary CFT in two dimensions.
The argument uses an assumption beyond unitarity on the removal of the regulator at the endpoints of the interval, which is supported by vacuum dominance in the closed channel by a gapped spectrum of a compact unitary CFT. Its precise statement is given as Assumption~(\ref{ass:regulator}) in the \SM.

\section{Strong subadditivity implies $c_{LR}\le c_{\mathrm{eff}}$}

We show how strong subadditivity of $S(u)$ combined with its quadratic variation formula
\eqref{eq:curvature-target-expansion} 
implies the desired inequality. 

Let us move the endpoints inward along future-directed null rays:
\[
  P_1(p)=(-a+p/2,p/2),
  \qquad
  P_2(q)=(b-q/2,q/2),
\]
whose cross-ratio is
\[
  \chi(p,q)
  =\frac{(a+b-p)(a+b-q)}{(2a-p)(2b-q)}.
\]
Denote the entropy of the
interval with these endpoints by $S(p,q)$. 
In the folded picture, the endpoint factors depend only on the distances $a-p/2$ and $b-q/2$ of the endpoints from the interface, so
\begin{equation}
  S(p,q)=\frac{c_L}{6}\log\frac{2a-p}{\epsilon}+\frac{c_R}{6}\log\frac{2b-q}{\epsilon}+F(u),
  \label{eq:boosted-entropy}
\end{equation}
with $\cosh^2u=\chi(p,q)$.
This reduces to \eqref{eq:entropy-u} at $p=q=0$, and for the trivial interface it reproduces $\frac{c}{6}\log\bigl[(\Delta x^2-\Delta t^2)/\epsilon^2\bigr]$. We choose the shifts so that all the intervals we consider below are spacelike, cross the interface, and have $\chi\ge1$. In this case, all insertions and their images in the folded picture are mutually spacelike (and null at $\chi=1$), as for static intervals, so the Lorentzian correlator equals the Euclidean one at the same cross-ratio and the same function $F$ applies. Moreover, $\sigma_L$ and $\sigma_R$ act on different tensor factors of the folded theory, so for every replica number the correlator is analytic at $\chi=1$. Hence $F$ is analytic and even near $u=0$.

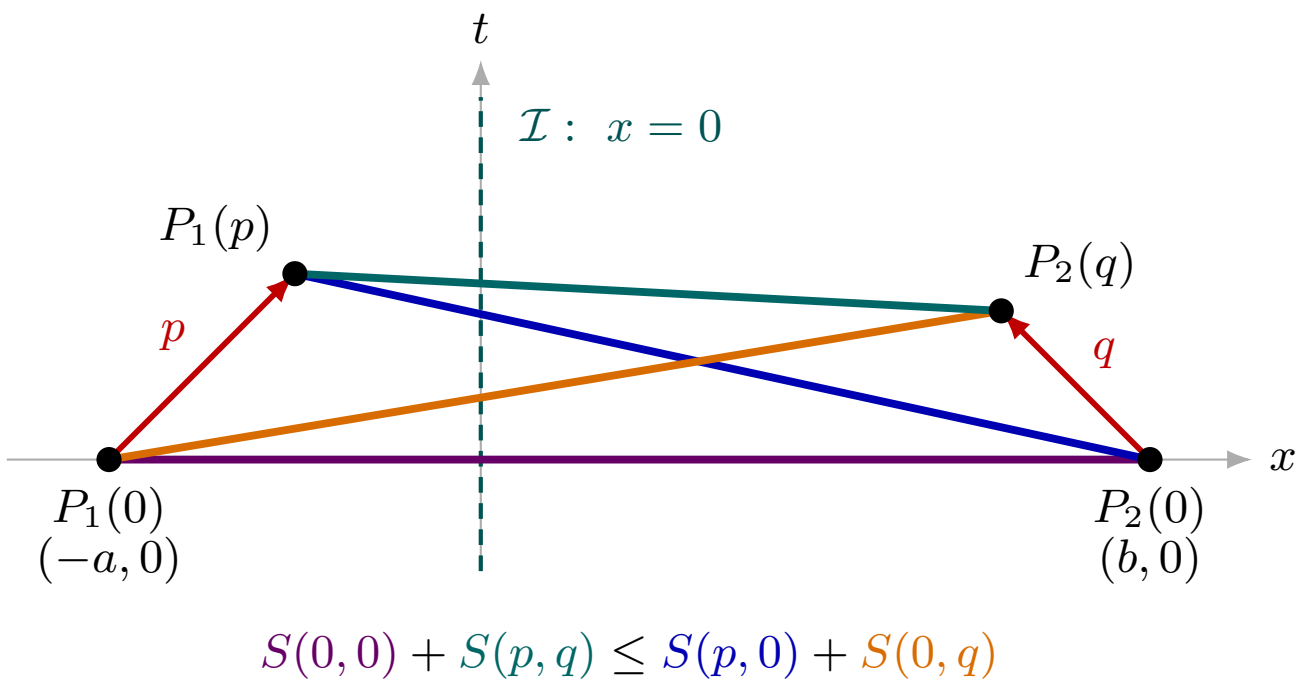
\begin{figure}[t]
\centering
\resizebox{\linewidth}{!}{%
\begin{tikzpicture}[
  x=1.25cm,y=1.25cm,
  >=Latex,
  font=\small,
  point/.style={
    circle,fill=black,inner sep=1.7pt
  },
  interval/.style={
    line width=1.5pt
  },
  null/.style={
    red!75!black,line width=1.1pt,
    -{Latex[length=2mm]}
  },
  entropy/.style={
    fill=white,inner sep=2pt,font=\small
  }
]

\colorlet{ssaA}{blue!70!black}
\colorlet{ssaB}{orange!85!black}
\colorlet{ssaU}{violet!80!black}
\colorlet{ssaI}{teal!80!black}

\coordinate (L0) at (-2,0);
\coordinate (Lp) at (-1,1);
\coordinate (Rq) at (2.8,.8);
\coordinate (R0) at (3.6,0);

\draw[->,gray!65]
  (-2.55,0)--(4.15,0)
  node[right,black] {$x$};

\draw[->,gray!65]
  (0,-.6)--(0,2.15)
  node[above,black] {$t$};

\draw[teal!65!black,densely dashed,line width=.8pt]
  (0,-.6)--(0,1.95);

\node[teal!65!black,anchor=west] at (.12,1.8)
  {$\mathcal I:\ x=0$};

\draw[interval,ssaU] (L0)--(R0); 
\draw[interval,ssaA] (Lp)--(R0); 
\draw[interval,ssaB] (L0)--(Rq); 
\draw[interval,ssaI] (Lp)--(Rq); 

\draw[null] (L0)--(Lp);
\draw[null] (R0)--(Rq);

\node[red!75!black,above left,inner sep=3pt]
  at ($(L0)!.5!(Lp)$) {$p$};

\node[red!75!black,above right,inner sep=3pt]
  at ($(R0)!.5!(Rq)$) {$q$};





\node[above left,inner sep=4pt]
  at (Lp) {$P_1(p)$};

\node[above right,inner sep=4pt]
  at (Rq) {$P_2(q)$};

\node[below,align=center,inner sep=5pt] at (L0)
  {$P_1(0)$\\[-1pt]$(-a,0)$};

\node[below,align=center,inner sep=5pt] at (R0)
  {$P_2(0)$\\[-1pt]$(b,0)$};

\foreach \P in {L0,Lp,Rq,R0}
  \node[point] at (\P) {};

\node[font=\small] at (.8,-1.05) {
  $
   \textcolor{ssaU}{S(0,0)}
   +\textcolor{ssaI}{S(p,q)}
   \le \textcolor{ssaA}{S(p,0)}
   +\textcolor{ssaB}{S(0,q)}$
};

\end{tikzpicture}%
}
\caption{
The four intervals entering strong subadditivity.
Red arrows indicate inward, future-directed null shifts
of the endpoints. Colored segments connect the endpoint
pairs defining the corresponding entropies.
}
\label{fig:SSA-null-intervals}
\end{figure}

As illustrated in Fig.~\ref{fig:SSA-null-intervals}, let $A$ run from $P_1(p)$ to $P_2(0)$ and $B$ from $P_1(0)$ to $P_2(q)$. The null segments $[P_1(0),P_1(p)]$ and $[P_2(q),P_2(0)]$ are limits of spacelike segments, so $A$ and $B$ lie on a common Cauchy curve. On this curve, $A\cup B$ runs from $P_1(0)$ to $P_2(0)$ and $A\cap B$ from $P_1(p)$ to $P_2(q)$. Strong subadditivity therefore gives \cite{Casini_2007}
\begin{equation}
     S(0,0)+S(p,q) \leq   S(p,0)+S(0,q).
\end{equation}
Each endpoint occurs once on either side, so the endpoint terms in \eqref{eq:boosted-entropy} cancel, leaving an inequality for $F$.

It is convenient to express the finite entropy in terms of
the logarithmic cross-ratio:
\[
  \Phi(\log\chi)  :=F\!\left(u\right)-F(0).
\]
Since $\chi(p,q)$ is a product of a function of $p$ and a function of $q$,
$\log\chi$ is additive in the two shifts. For $b>a$ and $0\le q\le b-a$, write
\begin{align*}
  X&:=\log\frac{(a+b)^2}{4ab},\\
  Y&:=\log\frac{2a(a+b-p)}{(a+b)(2a-p)},\\
  Z&:=\log\frac{(a+b)(2b-q)}{2b(a+b-q)},
\end{align*}
so that $Y\ge0$ and $0\le Z\le X$, with $Z=X$ at $q=b-a$.
Consequently, the four intervals have
\[
\begin{array}{c|cccc}
  \text{entropy}
    &S(p,0)&S(0,q)&S(0,0)&S(p,q)\\ \hline
  \log\chi
    &X+Y&X-Z&X&X+Y-Z
\end{array}
\]
and strong subadditivity implies
\begin{equation}
  \Phi(X)+\Phi(X+Y-Z)\le\Phi(X+Y)+\Phi(X-Z).
  \label{eq:convex}
\end{equation}
Namely, the increment $\Phi(X+Y)-\Phi(X)$ over a step $Y$ does not
decrease as $X$ increases, so that $\Phi(X)$ is convex,
$\Phi''(X)\ge0$ for $X\ge0$. Indeed, for infinitesimal $p$ and $q$,
\eqref{eq:convex} reads $\Phi''(X)\,YZ\ge0$.

On the other hand, as the interval becomes largely asymmetric and one end is much longer than the other, this corresponds to large $u$ where $X=\log\cosh^2u=2|u|+O(1)$, and the effective central charge is given by \cite{karch2023universalityeffectivecentralcharge,wang2026twistoperatorbopeentanglement}
\[
  c_{\mathrm{eff}}
  =3\lim_{u\to\infty}\frac{F(u)-F(0)}{u}
  =6\lim_{X\to\infty}\Phi'(X),
\]
where the limit of $\Phi'(X)$ exists since it is non-decreasing by the convexity.
Recall that \eqref{eq:curvature-target-expansion} relates the slope of $\Phi$ at the origin to the
transmission coefficient, $c_{LR}=6\Phi'(0)$, while the large-$X$
behavior gives $c_{\rm eff}=6\Phi'(\infty)$. Their difference can be
expressed as
\begin{equation}
    c_{\rm eff}-c_{LR}
    = 6\int_0^\infty \dd X\,\Phi''(X)
    \geq 0.
\end{equation}
The integrand is nonnegative by the convexity of $\Phi$. This establishes $c_{LR}\leq c_{\rm eff}$.

The bound is saturated by the trivial interface, for which $F(u)-F(0)=\frac{1}{6}c\log\cosh^2u$ and $c_{\mathrm{eff}}=c$, and by factorized interfaces, for which both sides vanish.

\section{Second derivative at symmetric interval}

We now derive \eqref{eq:curvature-target-expansion}, which relates the variation of the entanglement entropy (interface entropy to be precise) of the symmetric interval $[-\ell,\ell]$ when moving the endpoints to the energy transmission $c_{LR}$. We will use the integral expression for the density matrix of the replica method, and calculate the second-order derivative of the entanglement entropy using the conformal Ward identity.

Varying the places of the boundary points of the symmetric interval corresponds to inserting stress tensors in the action. To be concrete we Wick rotate to Euclidean signature via $y=it$ so that $z=x+iy$ and move the two endpoints
of $[-\ell,\ell]$ to $-a=-\ell-\delta a$ and $b=\ell+\delta b$. The first-order variation of the action is
$\delta I=\delta a\,V_L+\delta b\,V_R$, where $V_L$ and $V_R$ are integrals of the stress tensors of $\mathrm{CFT}_L$ and $\mathrm{CFT}_R$ over regions around the left and the right endpoints, respectively.

Let us now do a conformal map to the cylinder picture where the interval runs along the cylinder between the cutoff walls, which is convenient for performing the replica trick \cite{SakaiSatoh2008Entanglement,Cardy_2016}

\begin{equation}
 w=\log\frac{z+\ell}{\ell-z}=s+i\tau,\qquad
 0\leq\tau\leq2\pi.
 \label{eq:endpoint-reference-map}
\end{equation}
The original interface $x=0$ maps to $s=0$, the two endpoints are sent to $s\to\mp\infty$, and the upper and lower banks of the interval cut map to $\tau=0$ and $\tau=2\pi$. The interface thus becomes a straight line along the $\tau$ direction, and the cylinder with the interface is invariant under translations of $\tau$.
We also impose cutoff walls at $s=\pm W$, corresponding to the cutoffs $\epsilon=\ell/\sinh W$ in the original $z$-plane in \eqref{eq:entropy-u}. See Fig.~\ref{fig:endpoint-cylinder} in the \SM\ for the setup after the transformation.

Let $H$
generate translations in Euclidean time $\tau$ on the regulated
spatial segment $s\in[-W,W]$ in the interface vacuum state.
Evolution through one period gives the reference density matrix
\begin{equation} \rho_0=e^{-2\pi H}/Z.
\label{rho0}
\end{equation}
Here $Z=\Tr e^{-2\pi H}$ is the partition function on the regulated cylinder.
The modular Hamiltonian for the interval is therefore $K_0=-\log\rho_0
=2\pi H+\log Z$. In terms of the original coordinate $x$ on the interval, it is
\begin{equation}
 K_0=2\pi\int_{-\ell}^{\ell}\dd x\,\frac{\ell^2-x^2}{2\ell}\,T_{00}(x)+{\rm const.},
 \label{eq:modular-hamiltonian}
\end{equation}
where $T_{00}=-(T+\bar T)/(2\pi)$ is the energy density, of $\mathrm{CFT}_L$ for $x<0$ and of $\mathrm{CFT}_R$ for $x>0$.

Let us now calculate the deformation of entanglement entropy. Let $\rho(a,b)$ be the normalized density operator for the interval $[-a,b]$. We denote its value at the reference $a=b=\ell$ state $\rho_0$, its first-order derivatives on the endpoints at both ends 
by $\rho_a,\rho_b$, and the second-order derivative $\rho_{ab}$. Differentiating the normalized cut path integral with respect to $a$
inserts $-V_L$, and similarly for $b$. Write $K(a,b)=-\log\rho(a,b)$ and
$K_p=\left.\partial_p K\right|_{a=b=\ell}$ for $p=a,b$. From above we have $K_0\equiv K(\ell,\ell)$.
The change $\delta\rho$ in the density matrix induced by the endpoint
variation need not commute with the reference state $\rho_0$. To account
for this noncommutativity, we use an integral representation of the
operator logarithm
\cite{Faulkner_2016},
\[
 \log\rho(a,b)=\int_0^\infty\dd r\,
 \left[\frac{\mathbf 1}{1+r}-(\rho+r\mathbf 1)^{-1}\right].
\]
As we show in Sec.~\ref{sm:setup} of the \SM, we can show that the second-order derivative of the entanglement entropy on both endpoints yield \cite{Faulkner_2016}
\[
 \left.\partial_b\partial_a S\right|_{a=b=\ell}
 =\Tr(\rho_{ab}K_0)+\Tr(\rho_a K_b).
\]
The first trace is the mixed variation of the expectation value of
the fixed modular Hamiltonian $K_0$. The second pairs the two
first-order endpoint variations through the resolvent. In terms of the
Bogoliubov--Kubo--Mori (BKM) metric \cite{PetzToth1993}, we obtain
\begin{equation}
 \left.\partial_a\partial_b S(a,b)\right|_{a=b=\ell}
 =\mathcal C_{ab}-g_{\BKM}(\rho_a,\rho_b),
 \label{eq:entropy-Hessian-fixed-W}
\end{equation}
where $\mathcal C_{ab}=\Tr(\rho_{ab}K_0)$ and
\begin{align}
 &g_{\BKM}(\rho_a,\rho_b)
 \equiv-\Tr(\rho_a K_b)\nonumber\\
 &\quad=\int_0^\infty\dd r\,
 \Tr\!\left[\rho_a(\rho_0+r\mathbf 1)^{-1}
            \rho_b(\rho_0+r\mathbf 1)^{-1}\right].
 \label{eq:BKM-definition}
\end{align}
In terms of the cross-ratio $u$ defined above, we have $\left.\partial_a\partial_b S(a,b)\right|_{a=b=\ell}
 =-S''(0)/4\ell^2=-F''(0)/4\ell^2$.

Both terms in \eqref{eq:entropy-Hessian-fixed-W} are given by correlation functions of the stress tensors on the cylinder: the BKM metric by the two-point function of $T_L$ and $T_R$, and the modular energy $\mathcal C_{ab}$ by their three-point function with the stress tensor in $K_0$. Since the interface is straight on the cylinder, the map $z=\ell\tanh(w/2)$ brings the connected part of the 2-pt function of stress tensors in the state $\rho_0$ to \cite{Billo2016Defects}
\begin{align}
 \mathcal G(w_{12})
 &:=\big\langle T_L(w_1)T_R(w_2)\big\rangle_{0,c}\nonumber\\
 &\phantom{:}=\frac{\clr}{2[2\sinh(w_{12}/2)]^4},
 \label{eq:endpoint-cyl-correlators}
\end{align}
with $w_{12}=w_1-w_2$, and the connected 3-pt functions are also fixed by $c_{LR}$ \cite{Billo2016Defects,Meineri:2019ycm}. Using them, we show in Secs.~\ref{sec:calculation} and \ref{sec:modular-energy} of the \SM\ that
\begin{equation}
 g_{\BKM}(\rho_a,\rho_b)=\frac{\clr}{12\ell^2},\qquad
 \mathcal C_{ab}=0,
 \label{eq:main-results}
\end{equation}
after removing the cutoff walls by Assumption~(A). In the BKM metric, the integral over the frequency conjugate to the modular time is given by a single pole, and the result does not depend on the positions of the two insertions. The modular energy vanishes because the insertions which move the endpoints contain only the harmonics $1$ and $e^{\pm i\tau}$ along the $\tau$ circle, while the response of $\mathcal G$ to $K_0$ contains only $e^{\pm im\tau}$ with $m\geq2$. Together with \eqref{eq:entropy-Hessian-fixed-W}, we obtain
\begin{equation}
 \left.\partial_a\partial_b S(a,b)\right|_{a=b=\ell}
 =-\frac{\clr}{12\ell^2},\qquad F''(0)=\frac{\clr}{3},
 \label{eq:endpoint-mixed-result}
\end{equation}
which proves \eqref{eq:curvature-target-expansion}.

\section{Discussion}

We have shown that $c_{LR}\le c_{\mathrm{eff}}$ for conformal interfaces between compact unitary CFTs in two dimensions. It implies that in general conformal interfaces allow more entanglement transmission than energy transmission. Together with the complementary inequality $c_{\mathrm{eff}}\le\min(c_L,c_R)$
proven in \cite{karch2023universalityeffectivecentralcharge}, we have $c_{LR}\le c_{\mathrm{eff}}\le\min(c_L,c_R)$.
The proof combines two inputs of different nature: Strong subadditivity gives the global constraint \eqref{eq:convex}, the convexity of the entropy as a function of the logarithmic cross-ratio; the curvature of that same function is identified at the symmetric point with the energy transmission $c_{LR}$.

Let us explain the idea behind the second input \eqref{eq:curvature-target-expansion}. Moving an endpoint deforms the entangling region, and such a deformation is generated by the stress tensor. In a 2D CFT without an interface, a single insertion of the stress tensor gives the dependence of the entropy on the length of the interval, which is fixed by the central charge \cite{Calabrese_2004}. In general, the response of the entanglement entropy to deformations of a planar or spherical entangling surface has been studied extensively \cite{RosenhausSmolkin2014Perturbative,AllaisMezei2015Shape,Mezei2015DeformedSphere}. The second-order response, known as the entanglement density \cite{BhattacharyaEtAl2015EntanglementDensity,Faulkner_2016}, comes from two insertions of the stress tensor, and its universal part is proportional to the coefficient $C_T$ of the stress tensor two-point function \cite{Mezei2015DeformedSphere,Faulkner_2016,BuenoMyersWitczakKrempa2015Corner,BuenoWitczakKrempa2016Bounds}. Planar and spherical surfaces are special since their modular Hamiltonians are local \cite{CasiniHuertaMyers2011Derivation}, so that the second-order response reduces to a two-point function in the vacuum.

Our relation \eqref{eq:curvature-target-expansion} is the counterpart of this result in the presence of a conformal interface (not to be confused with the entangling defect, which in our case is the endpoint of the symmetric interval across the physical interface), with $C_T$ replaced by the coefficient $c_{LR}$ of the stress tensor two-point function across the interface. For this, it is essential that the reference interval is symmetric about the interface. As we explain in Sec.~\ref{sm:symmetric} of the \SM, only for the symmetric interval does the modular flow map the interface to itself, so that the modular Hamiltonian remains local and is given by \eqref{eq:modular-hamiltonian}.


Away from the symmetric interval, $F(u)$ depends on the details of the interface, and we can calculate it exactly from the twist fields correlator in the replica method. In Sec.~\ref{app:bope} of the \SM, we expand their two-point function in two ways: by the boundary operator product expansion (BOPE) of the twist fields \cite{wang2026twistoperatorbopeentanglement}, where $c_{\mathrm{eff}}$ is the dimension of the leading operator, and by their bulk OPE in the folded picture, where $c_{LR}$ appears at the first subleading order. Comparing the two, we express $c_{\mathrm{eff}}-c_{LR}$ in terms of the BOPE data in \eqref{eq:bope-difference}, so that our bound becomes a constraint on the (B)OPE coefficients.

The bound $c_{LR}\le c_\text{eff}$ is saturated by both the trivial and the factorized interfaces. From the argument above, the saturation is equivalent to $\Phi(X)$ being linear, which fixes the form of entanglement entropy of intervals in the vacuum interface state at any position $[-a,b]$. In the language of Sec.~\ref{app:bope}, it is equivalent to fixing the large-asymmetric interval boundary entropy $F_*$ to be equal to the standard interface entropy $\log g$ \footnote{In terms of the large-asymmetric interface entropy $\log \tilde{g}^{(b)}$ in \cite{wang2026twistoperatorbopeentanglement}, the saturation condition to this bound is equivalent to saturation of the inequality $\log g-\log \tilde{g}^{(b)}\ge \frac{c_\text{eff}}{3}\log 2$.}. Unlike the holographic case in \cite{karch2024universalbound}, for generic conformal interfaces it is not clear what are the necessary conditions on the interface and ambient CFTs for this saturation. For example, there can be interfaces that are transparent on one tensor factor and factorizing on the others, with $c_{LR}=c_{\mathrm{eff}}$ equal to the central charge of the transmitted factor \cite{harris2025holographicinterfacessymmetricproduct}. Coset constructions also allow transmission of a common conformal subalgebra while reflecting its complement, without a tensor-product decomposition of the full CFT \cite{quella2007reflection,Meineri:2019ycm,karch2024universalbound}. For certain systems, equality in strong subadditivity is characterized by quantum Markov structure \cite{Hayden_2004,CasiniTesteTorroba2017Markov,FawziRenner2015Recovery}. Studying what this means for the continuum null configurations considered here, and fully classifying all conformal interfaces that saturate this bound are left for future work. In addition, one may also study new bounds between interface observables, such as the one in \cite{kusuki2026entanglementrequiresfluctuationsconformal} between $c_\text{eff}$ and the capacity of entanglement $c_{\text{cap}}$ using similar techniques.

\section*{Acknowledgments}
We would like to thank Marco Meineri for his participation in an earlier part of this project and for his discussions. We would like to thank Yuya Kusuki and Hao-Yu Sun for useful discussions. 
GPT-6 Astra made significant contributions to the evaluation of the second-order variations of the entanglement entropy. Claude Opus helped us refine an earlier version of the modular-energy argument.

AK is supported in part by the U.S. Department of Energy grant DE-SC0022021 and by a grant from the Simons Foundation (Grant 651678, AK). HO is supported in part by the U.S. Department of Energy, Office of Science, Office of High Energy Physics, under Award Number DE-SC0011632,
the Simons Investigator Award (MP-SIP-00005259), and
JSPS Grants-in-Aid for Scientific Research 23K03379. 
His work was performed in part at Pioneering Science Promotion Division of RIKEN. MW is supported by World Premier International Research Initiative (WPI), MEXT, Japan.

\bibliography{references}

\clearpage
\onecolumngrid
\begin{center}
\textbf{\large Supplemental Material}
\end{center}
\setcounter{equation}{0}
\setcounter{figure}{0}
\setcounter{section}{0}
\setcounter{secnumdepth}{2}
\renewcommand{\theequation}{S\arabic{equation}}
\renewcommand{\thefigure}{S\arabic{figure}}
\renewcommand{\thesection}{\Alph{section}}
\renewcommand*{\theHfigure}{supp.\arabic{figure}}
\renewcommand*{\theHequation}{supp.\arabic{equation}}

In this Supplemental Material, we give the details of the derivation of \eqref{eq:curvature-target-expansion} and discuss the idea behind it. In particular, in Sec.~\ref{sm:setup} we describe the endpoint variations on the cylinder and derive the formula \eqref{eq:entropy-Hessian-fixed-W} for the mixed second derivative of the entropy. In Secs.~\ref{sec:calculation} and \ref{sec:modular-energy} we evaluate the BKM metric and the modular energy in \eqref{eq:main-results}. In Sec.~\ref{sm:symmetric} we write down the modular Hamiltonian in the presence of the interface, and discuss why the symmetric interval is special and what happens for asymmetric intervals. Finally in Sec.~\ref{app:bope} we take a step further and express $c_{\mathrm{eff}}-c_{LR}$ directly in terms of the (boundary) operator product expansion of the 2-pt function of twist fields in the replica method. The strong subadditivity of entanglement entropy then puts a constraint on their (B)OPE coefficients.

\section{The second-order response of entanglement entropy}
\label{sm:setup}

Let us now calculate the second-order derivative of the entanglement entropy of the symmetric interval in \eqref{eq:curvature-target-expansion} when moving its endpoints, which corresponds to inserting stress tensors in the action. To be concrete we Wick rotate to Euclidean signature via $y=it$ so that $z=x+iy$ and move the two endpoints
of $[-\ell,\ell]$ to $-a=-\ell-\delta a$ and $b=\ell+\delta b$ by the change of coordinates
$z\mapsto z+\delta a\,\xi_L+\delta b\,\xi_R$.
Here $\xi_L$ and $\xi_R$ are real profiles, with $\xi_L=-1$ near the left endpoint and $\xi_R=1$ near the
right endpoint. Each of them drops to zero before reaching the interface, so that the endpoints move while the interface stays at $x=0$. The action changes only where the profiles vary, and its first-order variation is
$\delta I=\delta a\,V_L+\delta b\,V_R$, where \cite{Meineri:2019ycm}
\begin{equation}
 V_\alpha=\frac1\pi\int\dd^2z\,
 \big[(\partial_{\bar z}\xi_\alpha)T_\alpha(z)
    +(\partial_z\xi_\alpha)\bar T_\alpha(\bar z)\big].
 \label{eq:endpoint-plane-insertion}
\end{equation}
Here $\alpha=L,R$ are the left and right half-planes and
$T_\alpha=-2\pi T^\alpha_{zz}$.

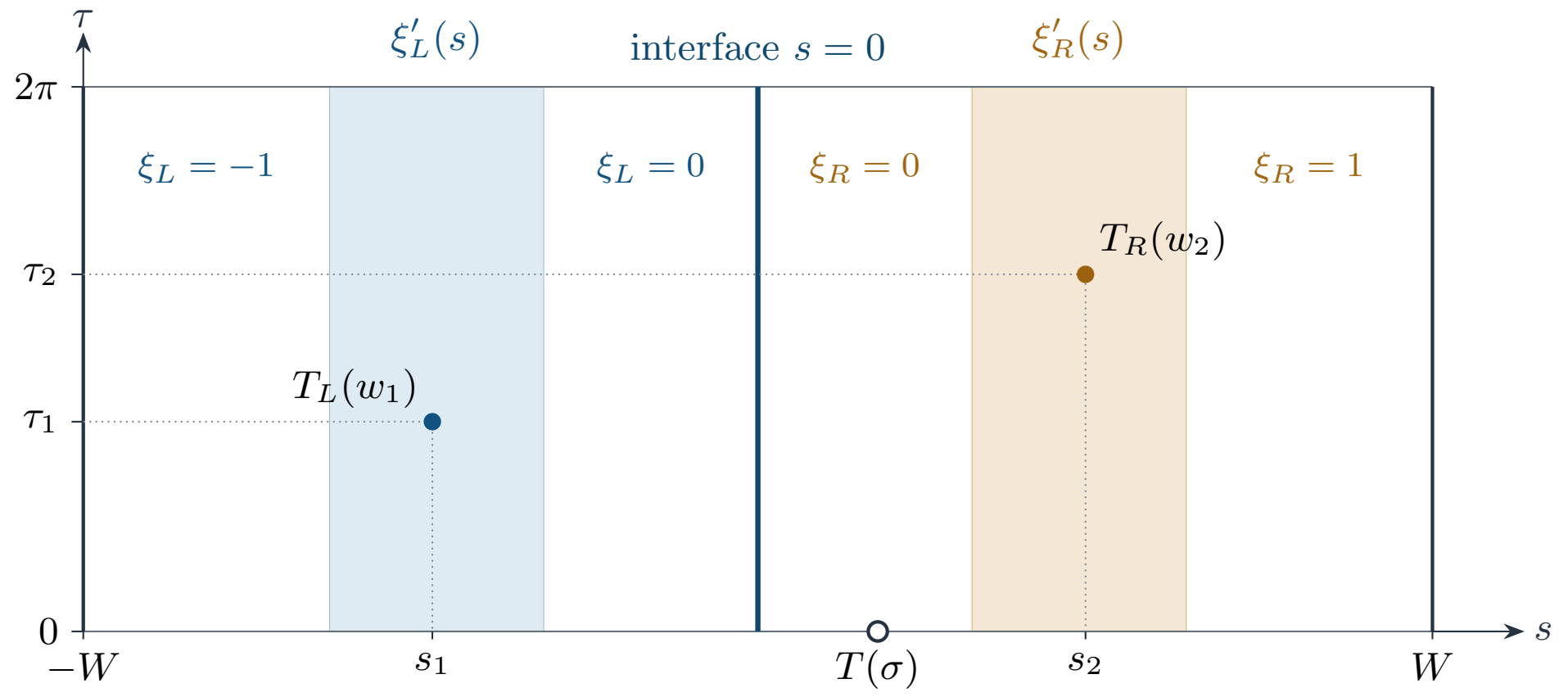
\begin{figure}[!htb]
 \centering
 \begin{tikzpicture}[x=1.73cm,y=.70cm,>=Stealth,font=\small]
  \def\wall{3.15}
  \pgfmathsetmacro{\tautop}{2*pi}
  \coordinate (leftsource) at (-1.52,2.42);
  \coordinate (rightsource) at (1.53,4.12);
  \coordinate (charge) at (.56,0);

  \fill[defect!14] (-2,0) rectangle (-1,\tautop);
  \fill[banklower!18] (1,0) rectangle (2,\tautop);
  \foreach \s in {-2,-1}
     \draw[defect!45,very thin] (\s,0)--(\s,\tautop);
  \foreach \s in {1,2}
     \draw[banklower!58,very thin] (\s,0)--(\s,\tautop);

  \draw[ink!70] (-\wall,0)--(\wall,0)
                  (-\wall,\tautop)--(\wall,\tautop);
  \draw[ink,thick] (-\wall,0)--(-\wall,\tautop)
                    (\wall,0)--(\wall,\tautop);
  \draw[->,ink] (-\wall,\tautop)--(-\wall,{\tautop+.63})
       node[above,inner sep=1pt] {$\tau$};
  \draw[->,ink] (\wall,0)--({\wall+.43},0)
       node[right,inner sep=2pt] {$s$};
  \draw[defect!65!black,very thick] (0,0)--(0,\tautop);
  \node[above=3pt,defect!65!black] at (0,\tautop) {interface $s=0$};
  \node[above=3pt,defect!75!black] at (-1.50,\tautop) {$\xi_L'(s)$};
  \node[above=3pt,banklower!82!black] at (1.50,\tautop) {$\xi_R'(s)$};
  \node[defect!75!black,font=\footnotesize] at (-2.575,5.35) {$\xi_L=-1$};
  \node[defect!75!black,font=\footnotesize] at (-.5,5.35) {$\xi_L=0$};
  \node[banklower!82!black,font=\footnotesize] at (.5,5.35) {$\xi_R=0$};
  \node[banklower!82!black,font=\footnotesize] at (2.575,5.35) {$\xi_R=1$};

  \draw[ink!55,densely dotted] (-\wall,2.42)--(leftsource)--(-1.52,0);
  \draw[ink!55,densely dotted] (-\wall,4.12)--(rightsource)--(1.53,0);
  \foreach \y/\lab in {0/0,2.42/\tau_1,4.12/\tau_2,\tautop/{2\pi}} {
     \draw[ink] (-\wall,\y)--({-\wall-.055},\y);
     \node[anchor=east,inner sep=2pt] at ({-\wall-.07},\y) {$\lab$};
  }
  \foreach \s/\lab in {{-\wall}/{-W},-1.52/s_1,1.53/s_2,\wall/W} {
     \draw[ink] (\s,0)--(\s,-.09);
     \node[below=2pt,inner sep=1pt] at (\s,-.09) {$\lab$};
  }

  \fill[defect!75!black] (leftsource) circle[radius=2pt];
  \node[above left=1pt,inner sep=2pt] at (leftsource) {$T_L(w_1)$};
  \fill[banklower!80!black] (rightsource) circle[radius=2pt];
  \node[above right=1pt,inner sep=2pt] at (rightsource) {$T_R(w_2)$};
  \draw[ink,fill=white,line width=.8pt] (charge) circle[radius=2.2pt];
  \node[below=3pt,inner sep=1pt] at (charge) {$T(\sigma)$};
 \end{tikzpicture}
 \caption{Reference cylinder with spatial walls at $s=\pm W$.
 The two banks of the cut, $\tau=0$ and $\tau=2\pi$, are identified in a trace.
 The profiles $\xi_L$ and $\xi_R$ vary only inside the shaded bands. The stress
 tensors $T_L(w_1)$ and $T_R(w_2)$ (filled points) are inserted there and
 integrated over the bands, and $\bar T_L$, $\bar T_R$ are inserted at the same points.
 The open circle is the stress tensor $T(\sigma)$ on the cut used in Sec.~\ref{sec:modular-energy}.}
 \label{fig:endpoint-cylinder}
\end{figure}

We use the conformal map \eqref{eq:endpoint-reference-map} to the cylinder with the coordinate $w=s+i\tau$; see Fig.~\ref{fig:endpoint-cylinder} for the setup after the transformation. The original interface $x=0$ maps to the straight line $s=0$, and we impose cutoff walls at $s=\pm W$, corresponding to the cutoffs $\epsilon=\ell/\sinh W$ in the original $z$-plane in \eqref{eq:entropy-u}.

On the cylinder we can freely choose the profiles $\xi_L$ and $\xi_R$ to depend only on $s$. We let $\xi_L(s)$ increase from $-1$ to $0$ somewhere on the left of the interface, $s<0$, and $\xi_R(s)$ increase from $0$ to $1$ somewhere on the right, $s>0$, as shown by the two shaded bands in Fig.~\ref{fig:endpoint-cylinder}. The stress tensors are then inserted only inside these bands, with the weights $\xi_\alpha'(s)=\dd\xi_\alpha/\dd s$ normalized as
$\int\dd s\,\xi_\alpha'(s)=1$. As we will see, the result does not depend on any other detail of the profiles.

Since $\dd w/\dd z=(1+\cosh w)/\ell$, the vector field $\xi_\alpha\partial_z$ becomes
$\xi_\alpha(s)\,\ell^{-1}(1+\cosh w)\,\partial_w$ on the cylinder, and $\partial_{\bar z}\xi_\alpha$ in \eqref{eq:endpoint-plane-insertion} is replaced by
$\partial_{\bar w}[\xi_\alpha(s)(1+\cosh w)/\ell]=\xi_\alpha'(s)(1+\cosh w)/(2\ell)$.
The stress tensor insertion on each half of the cylinder is therefore
\begin{equation}
 V_\alpha=\frac1{2\pi\ell}\int\dd s\,\xi_\alpha'(s)
 \int_0^{2\pi}\dd\tau\,
 \big[(1+\cosh w)\,T_\alpha(w)
    +(1+\cosh\bar w)\,\bar T_\alpha(\bar w)\big]
 \label{eq:endpoint-sources}
\end{equation}
with $w=s+i\tau$, up to the Schwarzian term, a $c$-number which drops out after normalizing the density matrix.
The variation of the endpoints thus corresponds to stress tensor insertions at
$w_1=s_1+i\tau_1$ on the left and $w_2=s_2+i\tau_2$ on the right. Since $s_1<0<s_2$, the two insertions are always separated by the interface and never collide with each other, so we can integrate $\tau_1$ and $\tau_2$ over the full period.

The reference density matrix is $\rho_0=e^{-2\pi H}/Z$ in \eqref{rho0}, where $H$ generates translations in $\tau$. Under the evolution of $\tau$,
\begin{equation}
 T_\alpha(s+i\tau)=e^{\tau H}T_\alpha(s)e^{-\tau H},\qquad
 \bar T_\alpha(s-i\tau)=e^{\tau H}\bar T_\alpha(s)e^{-\tau H}.
 \label{eq:chiral-source-conventions}
\end{equation}
Differentiating the normalized cut path integral with respect to $a$
inserts $-V_L$. The two segments of Euclidean evolution, of lengths
$2\pi-\tau$ and $\tau$, give
\begin{equation}
\begin{aligned}
 \rho_a=-\frac{1}{2\pi\ell}\int\dd s\,\xi_L'(s)
 \int_0^{2\pi}\dd\tau\,
 \rho_0^{\,1-\tau/(2\pi)}
 \big[&(1+\cosh(s+i\tau))\,\delta T_L(s)\\
 &+(1+\cosh(s-i\tau))\,\delta\bar T_L(s)\big]
 \rho_0^{\,\tau/(2\pi)} .
\end{aligned}
 \label{eq:endpoint-state-variation}
\end{equation}
For $\rho_b$ we replace the left stress by the right stress and
$\xi_L'$ by $\xi_R'$. Here $\delta T_\alpha=T_\alpha-
\langle T_\alpha\rangle_0$, where $\langle\cdots\rangle_0$ is the expectation value in $\rho_0$, and likewise for $\bar T_\alpha$.
The one-point subtraction comes from differentiating the normalization.

Let us next derive the formula \eqref{eq:entropy-Hessian-fixed-W} for the mixed second derivative of the entropy, starting from the integral representation of the operator logarithm given in the main text.
Differentiating the inverse and then evaluating at the symmetric
interval gives both modular-Hamiltonian variations,
\[
 K_p=-\int_0^\infty\dd r\,
       (\rho_0+r\mathbf 1)^{-1}\rho_p(\rho_0+r\mathbf 1)^{-1},
 \qquad p=a,b .
\]
At general $a,b$, cyclicity of the trace gives
\[
 \Tr(\rho\,\partial_a K)
 =-\Tr\!\left[(\partial_a\rho)\int_0^\infty\dd r\,
                      \rho(\rho+r\mathbf 1)^{-2}\right]
 =-\partial_a\Tr\rho=0 .
\]
The integral is the identity on the support of $\rho$, since it acts
on each eigenvalue $\lambda>0$ as
$\int_0^\infty\lambda\,\dd r/(\lambda+r)^2=1$.

It follows that $\partial_a S=\Tr[(\partial_a\rho)K]$. Taking the
$b$ derivative and then setting $a=b=\ell$ gives
\[
 \left.\partial_b\partial_a S\right|_{a=b=\ell}
 =\Tr(\rho_{ab}K_0)+\Tr(\rho_a K_b).
\]
This is the formula used in the main text to obtain \eqref{eq:entropy-Hessian-fixed-W} with the BKM metric \eqref{eq:BKM-definition}. Cyclicity in \eqref{eq:BKM-definition} gives $\Tr(\rho_a K_b)=\Tr(\rho_b K_a)$.

Finally, let us state the assumption on the regulator mentioned in the main text.
\begin{itemize}
  \item[(A)]
  In the regularization of this section, removing the
  regulator walls ($W\to\infty$) commutes with the mixed endpoint derivative
  $\partial_a\partial_b$ at $a=b=\ell$, and the limits of both terms in
  \eqref{eq:entropy-Hessian-fixed-W} are given by correlators on the
  infinite cylinder. This is supported by vacuum dominance in the
  closed channel by a gapped spectrum of a compact unitary CFT.
  \label{ass:regulator}
\end{itemize}
For the vacuum projection we also assume a discrete defect spectrum with a gap above the vacuum sector.
Note that to derive the regulated expressions for the above quantities we calculate everything with the fixed spatial cutoff scheme $W$. Eventually we need to use Assumption~(A) to take $W\to\infty$
in \eqref{eq:entropy-Hessian-fixed-W}. It allows the wall limit to
pass through $\partial_a\partial_b$ and identifies the limit of each
term with its value on the infinite cylinder in the fixed interface
vacuum. The vacuum sector is selected by radial propagation
in the folded theory. Between semicircles of radii $r<R$, a defect
state of dimension $\widehat\Delta$ is suppressed relative to the
vacuum by $(r/R)^{\widehat\Delta}$. With the assumed defect gap and
normalizable outer data of nonzero vacuum overlap, the contribution
orthogonal to the vacuum sector vanishes as $R/r\to\infty$.

\section{The BKM metric}
\label{sec:calculation}

Let us first consider the BKM metric. In the energy eigenbasis $H|i\rangle=E_i|i\rangle$ of the reference Hamiltonian, where $\rho_0$ in \eqref{rho0} has the eigenvalues $p_i=e^{-2\pi E_i}/Z$, the resolvent integral \eqref{eq:BKM-definition}
gives the spectral sum \cite{PetzToth1993}
\begin{equation}
 g_{\BKM}(\rho_a,\rho_b)
 =\sum_{i,j}(\rho_a)_{ji}(\rho_b)_{ij}
       \frac{\log p_i-\log p_j}{p_i-p_j}.
 \label{eq:BKM-spectral}
\end{equation}
To evaluate the above equation, recall from \eqref{eq:endpoint-sources} that the stress tensor is inserted at the Euclidean time $\tau_1$ in the path integral over one period, so it has the factor $e^{-(2\pi-\tau_1)H}$ on its left and $e^{-\tau_1H}$ on its right. For the holomorphic part, this gives
\begin{equation}
 (\rho_a)_{ji}=-\frac{1}{2\pi\ell\,Z}\int\dd s_1\,\xi_L'(s_1)\int_0^{2\pi}\dd\tau_1\,(1+\cosh w_1)\,
 e^{-(2\pi-\tau_1)E_j}\,e^{-\tau_1E_i}\,\langle j|T_L(s_1)|i\rangle ,
 \label{eq:rho-a-matrix}
\end{equation}
with $w_1=s_1+i\tau_1$, and similarly for the anti-holomorphic part. Since $\rho(a,b)$ is normalized, the expectation values of the stress tensors are subtracted, and only the connected part contributes.

Let us now turn the sum over states into integral over the spectrum. Set $\omega=E_i-E_j$, the thermal weights are $p_ip_j\,e^{-\omega(\tau_1-\tau_2)}$. The last factor in \eqref{eq:BKM-spectral} is $-2\pi\omega/(p_i-p_j)$ with $p_i-p_j=-2\sqrt{p_ip_j}\,\sinh(\pi\omega)$, and therefore
\begin{equation}
 p_ip_j\,e^{-\omega(\tau_1-\tau_2)}\,\frac{\log p_i-\log p_j}{p_i-p_j}
 =\sqrt{p_ip_j}\;\frac{\pi\omega\,e^{-\omega(\tau_1-\tau_2)}}{\sinh(\pi\omega)} .
 \label{eq:thermal-weights}
\end{equation}
We can take the second factor out of the sum over states by inserting delta functions
\begin{equation}
 1=\int\dd\omega\,\delta(\omega-E_i+E_j)=\int\dd\omega\int\frac{\dd t}{2\pi}\,e^{i\omega t}\,e^{-i(E_i-E_j)t} .
\end{equation}
The remaining sum over states is
\begin{equation}
 \sum_{i,j}\sqrt{p_ip_j}\;e^{-i(E_i-E_j)t}\,\langle j|T_L(s_1)|i\rangle\langle i|T_R(s_2)|j\rangle
 =\Tr\!\big[\rho_0\,e^{\pi H}e^{iHt}\,T_L(s_1)\,e^{-iHt}e^{-\pi H}\,T_R(s_2)\big] ,
\end{equation}
where we used $\sqrt{p_ip_j}=e^{-\pi E_i}e^{-\pi E_j}/Z$. The phases $e^{\pm iHt}$ and $e^{\pm\pi H}$ shifts the stress tensor correlator into a thermal one $\langle T_L(s_1-t+i\pi)\,T_R(s_2)\rangle$ in the interface vacuum state. The same holds for the anti-holomorphic parts.

Up to this point the cutoffs are at finite $W$, and reflected stress tensors are contained in our correlators. We now remove the cutoff by Assumption~(\ref{ass:regulator}): the spectrum becomes continuous, and on the infinite cylinder the mixed-chirality
2-pt functions of stress tensors across the interface vanish. We therefore obtain
\begin{align}
 g_{\BKM}(\rho_a,\rho_b)
 ={}&\frac{2}{(2\pi\ell)^2}\operatorname{Re}
 \int\dd s_1\,\xi_L'(s_1)\int\dd s_2\,\xi_R'(s_2)
 \nonumber\\
 &\times\int_0^{2\pi}\dd\tau_1\int_0^{2\pi}\dd\tau_2\,
 (1+\cosh w_1)(1+\cosh w_2)
 \nonumber\\
 &\times\int\dd\omega\,
       \frac{\pi\omega\,e^{-\omega(\tau_1-\tau_2)}}{\sinh(\pi\omega)}
       \int\frac{\dd t}{2\pi}\,e^{i\omega t}
       \big\langle T_L(s_1-t+i\pi)T_R(s_2)\big\rangle_{0,c}.
 \label{eq:BKM-correlator-integral}
\end{align}

At $w_1=s_1-t+i\pi$ and $w_2=s_2$, the last integral in
\eqref{eq:BKM-correlator-integral} is
\begin{equation}
 \int\frac{\dd t}{2\pi}\,e^{i\omega t}\,
       \frac{\clr}{32\cosh^4((t+s_2-s_1)/2)}
 =\frac{\clr}{24}\,\frac{\omega(1+\omega^2)}{\sinh(\pi\omega)}\,
       e^{-i\omega(s_2-s_1)}.
 \label{eq:endpoint-stress-density}
\end{equation}
 All the factors of $\sinh(\pi\omega)$ then cancel, and we are left with
\begin{align}
 &g_{\BKM}(\rho_a,\rho_b)
 =\frac{\clr}{12\pi\ell^2}\operatorname{Re}
 \int\dd s_1\,\xi_L'(s_1)\int\dd s_2\,\xi_R'(s_2)
 \int\dd\omega\,\frac{e^{-i\omega(s_2-s_1)}}{1+\omega^2}
 \nonumber\\
 &\quad\times\big[1+\omega^2+\omega^2\cosh s_1+i\omega\sinh s_1\big]
        \big[1+\omega^2+\omega^2\cosh s_2-i\omega\sinh s_2\big].
 \label{eq:endpoint-full-BKM}
\end{align}
The seemingly divergent parts of the integral containing the polynomial of $\omega$ is only supported on $\delta(s_2-s_1)$ after integrating $\omega$, which vanishes since the two insertions are on opposite sides of the interface. The only contribution comes from the remaining term $[\cosh(s_2-s_1)+i\omega\sinh(s_2-s_1)]/(1+\omega^2)$. Its pole at $\omega=-i$ is the only contribution where the two brackets become $-e^{-s_1}$ and $-e^{s_2}$, so the $\omega$ integral gives $\pi$. The remaining integrals over $s_1$ and $s_2$ are just the normalization of the two variation profiles, and we obtain
\begin{equation}
 g_{\BKM}(\rho_a,\rho_b)=\frac{\clr}{12\ell^2}.
 \label{eq:endpoint-BKM}
\end{equation}

\section{The modular energy}
\label{sec:modular-energy}

We now calculate the modular-energy term
\[
 \mathcal C_{ab}=\Tr(\rho_{ab}K_0)
 =2\pi\Tr(\rho_{ab}H),
\]
since $\Tr\rho_{ab}=0$. The second order derivative $\rho_{ab}$ amounts to inserting two stress tensors
$V_L V_R$, and in addition we have the insertion of the modular Hamiltonian \eqref{eq:modular-hamiltonian}. On the cylinder,
after taking the wall limit, it is
$2\pi H=-\int\dd\sigma\,[T(\sigma)+\bar T(\sigma)]$.
The stress tensor lies on the cut at $(\sigma,0)$, marked by the open
circle in Fig.~\ref{fig:endpoint-cylinder}; $T(\sigma)$ is $T_L(\sigma)$ for $\sigma<0$ and
$T_R(\sigma)$ for $\sigma>0$. Substituting the endpoint insertions \eqref{eq:endpoint-sources} gives
\begin{align}
 \mathcal C_{ab}
 &=\big\langle V_LV_R K_0\big\rangle_{0,c}\nonumber\\
 &=-\frac{2}{(2\pi\ell)^2}\operatorname{Re}
 \int\dd s_1\,\xi_L'(s_1)\int\dd s_2\,\xi_R'(s_2)
 \int_0^{2\pi}\dd\tau_1\int_0^{2\pi}\dd\tau_2\,
 (1+\cosh w_1)(1+\cosh w_2)\nonumber\\
 &\quad\times\int\dd\sigma\,
 \big\langle T_L(w_1)T_R(w_2)T(\sigma)\big\rangle_{0,c},
 \label{eq:charge-correlator-integral}
\end{align}
where, as in \eqref{eq:BKM-correlator-integral}, the antiholomorphic stress tensors give the complex conjugate.

The connected 3-pt functions of holomorphic stress tensors across conformal interfaces are also fixed by $c_{LR}$ \cite{Billo2016Defects,Meineri:2019ycm}
\begin{equation}
 \begin{aligned}
 \big\langle T_L(w_1)T_R(w_2)T_\alpha(w_3)\big\rangle_{0,c}
 =\frac{\clr}{\prod_{i<j}[2\sinh(w_{ij}/2)]^2}.
 \end{aligned}
 \label{eq:endpoint-three-point}
\end{equation}
In terms of $x=e^\sigma$, the integrand is a rational function with double poles at $x_j=e^{w_j}$, and the integral gives
\begin{align}
 -\int\dd\sigma\,
  \big\langle T_L(w_1)T_R(w_2)T(\sigma)\big\rangle_{0,c}
 &=\big[4-2w_{12}\coth(w_{12}/2)\big]\mathcal G(w_{12})\nonumber\\
 &=(4+w_{12}\partial_{w_{12}})\mathcal G(w_{12}),
 \label{eq:endpoint-integrated-charge}
\end{align}
where the term linear in $w_{12}$ comes from $\log(-x_1)-\log(-x_2)=w_{12}$ on the principal branch.
This is the expected response to $K_0=2\pi H$, which shifts the period of the cylinder away from $2\pi$. Indeed, acting on the two-point function at period $\beta$ we find the same result,
\[
 -2\pi\partial_\beta\,\frac{\clr}{2}
 \left[\frac{\pi/\beta}{\sinh(\pi w_{12}/\beta)}\right]^4\bigg|_{\beta=2\pi}
 =(4+w_{12}\partial_{w_{12}})\mathcal G(w_{12}).
\]
The expression \eqref{eq:endpoint-integrated-charge} stays finite when an insertion approaches the cut, $\tau_j\to0$ or $2\pi$, since $\operatorname{Re}w_{12}=s_1-s_2<0$, so we do not to regularize the coincidental limits of $T(\sigma)$ and $T(w_i)$.

For $\operatorname{Re}w_{12}<0$, the two-point function expands as
\[
 \mathcal G(w_{12})=\frac{\clr}{12}
       \sum_{m=2}^\infty m(m^2-1)e^{mw_{12}} ,
\]
so that \eqref{eq:endpoint-integrated-charge} is a sum of the terms $(4+mw_{12})e^{mw_{12}}$ with $m\geq2$.
Since $w_{12}=s_1-s_2+i(\tau_1-\tau_2)$, each term gives in \eqref{eq:charge-correlator-integral} products of a $\tau_1$ integral with $e^{im\tau_1}$ and a $\tau_2$ integral with $e^{-im\tau_2}$, and the prefactor $4+mw_{12}$ adds a power of $\tau_j$ to at most one of them. Thus at least one of the two integrals is of the form
\begin{equation}
 \int_0^{2\pi}\dd\tau\,[1+\cosh(s+i\tau)]\,e^{\pm im\tau}=0,
 \qquad m=2,3,\ldots .
 \label{eq:endpoint-angular-zero}
\end{equation}
 Thus the modular energy term vanishes $\mathcal C_{ab}=0$. Together with
\eqref{eq:endpoint-BKM} and \eqref{eq:entropy-Hessian-fixed-W}, this gives \eqref{eq:endpoint-mixed-result} and proves \eqref{eq:curvature-target-expansion}.

\section{Modular Hamiltonian for symmetric and asymmetric intervals}
\label{sm:symmetric}

As emphasized in the main text, it is essential for our derivation of \eqref{eq:curvature-target-expansion} that the reference interval is symmetric about the interface. Here we explain why this is the case, write down the modular Hamiltonian in the presence of the interface, and show what goes wrong for asymmetric intervals.

In a CFT without an interface, the modular Hamiltonian of any interval $[-a,b]$ is local, since the modular flow is the conformal transformation which preserves the causal diamond of the interval. With the center $m=(b-a)/2$ and the radius $R=(a+b)/2$ of the interval, it is generated by the conformal Killing vector
\begin{equation}
 \xi=\frac{\pi}{R}\Big[\big(R^2-(x-m)^2-t^2\big)\,\partial_t-2t\,(x-m)\,\partial_x\Big],
 \label{eq:modular-flow}
\end{equation}
and the modular Hamiltonian is its charge on the interval at $t=0$ \cite{CasiniHuertaMyers2011Derivation,Cardy_2016},
\begin{equation}
 K_\xi=\int_{-a}^{b}\dd x\,\xi^t(0,x)\,T_{00}(x)
 =2\pi\int_{-a}^{b}\dd x\,\frac{(x+a)(b-x)}{a+b}\,T_{00}(x),
 \label{eq:K-xi}
\end{equation}
up to an additive constant. In the presence of the interface, this transformation remains a symmetry only if it maps the interface to itself. At the interface, the component of $\xi$ normal to it is
\begin{equation}
 \xi^x(t,0)=2\pi t\,\frac{b-a}{a+b},
 \label{eq:modular-flow-interface}
\end{equation}
which vanishes for all $t$ only when $a=b$.

\paragraph{Symmetric interval.}
For the symmetric interval, the modular flow is therefore still geometric. In the Euclidean cylinder picture in the main text, the interface is mapped to the straight line $s=0$. The path integral on the cylinder is then invariant under translations of $\tau$, and the density matrix is $\rho_0=e^{-2\pi H}/Z$. Here $H$ is the Hamiltonian of $\mathrm{CFT}_L$ on $s<0$ and $\mathrm{CFT}_R$ on $s>0$. In the original coordinate, the modular Hamiltonian is therefore \cite{Mintchev_2021}
\begin{equation}
 K_0=2\pi\int_{-\ell}^{0}\dd x\,\frac{\ell^2-x^2}{2\ell}\,T^L_{00}(x)
 +2\pi\int_{0}^{\ell}\dd x\,\frac{\ell^2-x^2}{2\ell}\,T^R_{00}(x)+\log Z,
 \label{eq:K0-interface}
\end{equation}
where $T^\alpha_{00}=-(T_\alpha+\bar T_\alpha)/(2\pi)$. This recovers \eqref{eq:modular-hamiltonian} which is also equal to \eqref{eq:K-xi} with $a=b=\ell$ \cite{CasiniHuertaMyers2011Derivation,Cardy_2016}. The same is true for a sphere centered on a planar conformal defect in higher dimensions \cite{JensenOBannon2013Defects}.

Note that the interface does not appear explicitly in \eqref{eq:K0-interface}. This is because the energy flux $T_L-\bar T_L=T_R-\bar T_R$ is continuous across the conformal interface, so that no energy is stored on it and $H$ is the sum of the energies of the two sides.

\paragraph{Asymmetric interval.}
On the other hand, if the interval is asymmetric with $a\neq b$, the flow generated by the Killing vector in \eqref{eq:modular-flow} pushes the interface sideways, and it is no longer a symmetry. The current $T_{\mu\nu}\xi^\nu$ is not conserved at the interface, where the stress tensor has the discontinuity
\begin{equation}
 D(t)=\big[T_L+\bar T_L-T_R-\bar T_R\big](t,0)
 =2\pi\big[T^R_{00}(t,0^+)-T^L_{00}(t,0^-)\big].
 \label{eq:displacement}
\end{equation}
This is the displacement operator of the interface \cite{Billo2016Defects,Meineri:2019ycm}. By the continuity of the energy flux, $D=2(T_L-T_R)$ at the interface, and \eqref{eq:TT-normalization} gives $\langle D(t)D(0)\rangle=C_D/t^4$ with $C_D=2(c_L+c_R-2c_{LR})$. The charge of $\xi$ on the slice at time $t$ changes at the rate $-\xi^x(t,0)D(t)/(2\pi)$, so that $K_\xi$ in \eqref{eq:K-xi} is not the modular Hamiltonian unless $a=b$ or $D=0$.

In the Euclidean picture, the map $w=\log\frac{z+a}{b-z}$ still sends the interval to the cylinder, but the interface $z=iy$ is now mapped to the curve
\begin{equation}
 s=\frac12\log\frac{a^2+y^2}{b^2+y^2},
 \label{eq:interface-curve}
\end{equation}
which moves from $s=\log(a/b)=-2u$ at $\tau=0$ to $s=0$ at $\tau=\pi$ and back. The evolution in $\tau$ is then generated by a $\tau$-dependent Hamiltonian on the curve $\delta s(\tau)=u\,(1-\cos\tau)+O(u^3)$:
\begin{equation}
 H(\tau)=H_\xi+\frac{\delta s(\tau)}{2\pi}\,\mathcal D+\cdots,
 \label{eq:H-tau}
\end{equation}
Here $H_\xi=K_\xi/(2\pi)$ is the Hamiltonian with the interface fixed at $s=-2u$, $\delta s(\tau)$ is the displacement of the interface from it, $\mathcal D$ is the displacement operator \eqref{eq:displacement} on the cylinder, and the dots are terms of higher order in $\delta s$. The density matrix is a $\tau$-ordered exponential,
\begin{equation}
 \rho(a,b)=\frac1Z\,\mathcal T\exp\Big[-\int_0^{2\pi}\dd\tau\,H(\tau)\Big],
 \label{eq:rho-ordered}
\end{equation}
instead of $e^{-2\pi H}$. Since $\mathcal D$ does not commute with $H_\xi$, the modular Hamiltonian is no longer a local integral of the stress tensor.

We can see this explicitly to the first order in the asymmetry. Expanding \eqref{eq:rho-ordered} in $\delta s$, we obtain
\begin{equation}
 \rho(a,b)=\rho_\xi-\frac{u}{2\pi}\int_0^{2\pi}\dd\tau\,(1-\cos\tau)\,
 \rho_\xi^{\,1-\tau/(2\pi)}\,\delta\mathcal D\,\rho_\xi^{\,\tau/(2\pi)}+O(u^2),
 \label{eq:rho-first-order}
\end{equation}
with $\rho_\xi=e^{-2\pi H_\xi}/Z$ and $\delta\mathcal D=\mathcal D-\langle\mathcal D\rangle$, which is of the same form as \eqref{eq:endpoint-state-variation}. The variation of the modular Hamiltonian $K=-\log\rho$ then follows from the resolvent formula in Sec.~\ref{sm:setup}. In the energy eigenbasis of Sec.~\ref{sec:calculation}, with $\omega=E_i-E_j$, it is
\begin{equation}
 (K-K_\xi)_{ij}=\frac{u}{2\pi}\,\delta\mathcal D_{ij}\int_0^{2\pi}\dd\tau\,(1-\cos\tau)\,
 \frac{\pi\omega\,e^{(\tau-\pi)\omega}}{\sinh(\pi\omega)}
 =\frac{u\,\delta\mathcal D_{ij}}{1+\omega^2}.
 \label{eq:K-first-order-spectral}
\end{equation}
As in \eqref{eq:endpoint-full-BKM}, the factors of $\sinh(\pi\omega)$ cancel. Since $1/(1+\omega^2)$ is the Fourier transform of $\frac12e^{-|\varsigma|}$, the right-hand side is the integral of $e^{i\varsigma H_\xi}\mathcal D\,e^{-i\varsigma H_\xi}$ over $\varsigma$ with this weight. To this order, it is $\mathcal D$ transported by the modular flow of the symmetric interval along the interface to the time $t=R\tanh(\varsigma/2)$, where it is equal to $(\dd t/\dd\varsigma)^2D(t)$ since the displacement operator has dimension two. In the original coordinates, we therefore obtain
\begin{equation}
 K(a,b)=2\pi\int_{-a}^{b}\dd x\,\frac{(x+a)(b-x)}{a+b}\,T_{00}(x)
 +\frac{b-a}{2(a+b)^2}\int_{-R}^{R}\dd t\,\big(R-|t|\big)^2D(t)
 +O\big((b-a)^2\big),
 \label{eq:K-asymmetric}
\end{equation}
up to an additive constant.

Notice that the second term is an integral of the displacement operator over the part of the interface inside the causal diamond of the interval, and not over the interval itself. To express $D(t)$ in terms of operators on the interval at $t=0$, we have to evolve the stress tensor through the interface, and the result is not local in general. 

As a sanity check we consider the two extreme cases. For a totally transmissive interface, $D=0$ and $K=K_\xi$ for any $a$ and $b$, as in a CFT without an interface. For a factorized interface, $T_\alpha=\bar T_\alpha$ at $x=0$, so that $T^\alpha_{00}(t,0)$ is given by the stress tensor on the interval at $t=0$, at the distance $|t|$ from the interface, and \eqref{eq:K-asymmetric} becomes the expansion to the first order in $b-a$ of
\begin{equation}
 2\pi\int_{-a}^{0}\dd x\,\frac{a^2-x^2}{2a}\,T^L_{00}(x)
 +2\pi\int_{0}^{b}\dd x\,\frac{b^2-x^2}{2b}\,T^R_{00}(x),
 \label{eq:K-factorized}
\end{equation}
which is exactly the sum of the modular Hamiltonians of the two boundary CFTs \cite{Cardy_2016}.

Equivalently, the non-symmetric interval is obtained from the symmetric one by displacing the interface by a finite amount, and the expansion of $F(u)$ around $u=0$ involves more data than $c_{LR}$ \cite{quella2007reflection,Billo2016Defects,Meineri:2019ycm}. As we will see in Sec.~\ref{app:bope}, $F(u)$ with finite $u$ depends on the full BOPE/OPE data of the interface and bulk CFTs.

\section{Twist field BOPE and $c_{\mathrm{eff}}-c_{LR}$}
\label{app:bope}

In the main text, $c_{LR}$ and $c_{\mathrm{eff}}$ appear as two limits of the same function $F(u)$ in the entanglement entropy of an interval $[-a,b]$: its curvature at the symmetric interval in \eqref{eq:curvature-target-expansion}, and its slope for highly asymmetric intervals. In this section we unify their expressions in terms of the twist field 2-pt function in the replica method, and express their difference directly by the operator product expansion (OPE) and the boundary operator product expansion (BOPE) of the twist fields studied in \cite{wang2026twistoperatorbopeentanglement}.

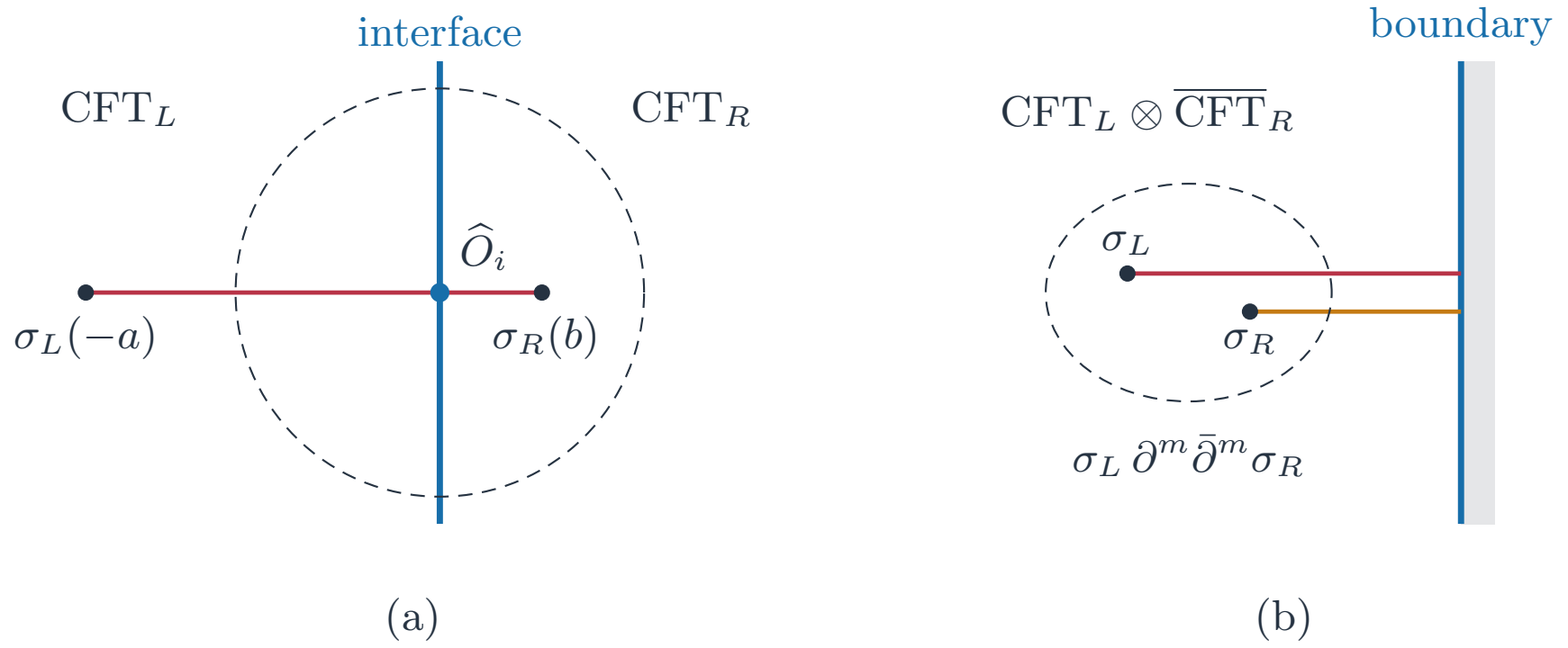
\begin{figure}[!htb]
\centering
\begin{tikzpicture}[font=\small]
\begin{scope}
 \draw[defect,line width=1.3pt] (0,-1.7)--(0,1.7);
 \node[defect,above] at (0,1.7) {interface};
 \node[ink] at (-2.35,1.35) {$\mathrm{CFT}_L$};
 \node[ink] at (1.85,1.35) {$\mathrm{CFT}_R$};
 \draw[bankupper,line width=0.9pt] (-2.6,0)--(0.75,0);
 \draw[ink,densely dashed] (0,0) circle (1.5);
 \fill[ink] (-2.6,0) circle (1.7pt);
 \fill[ink] (0.75,0) circle (1.7pt);
 \node[ink,below] at (-2.6,-0.06) {$\sigma_L(-a)$};
 \node[ink,below] at (0.78,-0.06) {$\sigma_R(b)$};
 \fill[defect] (0,0) circle (2pt);
 \node[ink,anchor=south west] at (0.03,0.05) {$\widehat O_i$};
 \node[ink] at (-0.2,-2.4) {(a)};
\end{scope}
\begin{scope}[xshift=7.5cm]
 \fill[ink!12] (0,-1.7) rectangle (0.25,1.7);
 \draw[defect,line width=1.3pt] (0,-1.7)--(0,1.7);
 \node[defect,above] at (0,1.7) {boundary};
 \node[ink] at (-2.3,1.35) {$\mathrm{CFT}_L\otimes\overline{\mathrm{CFT}}_R$};
 \draw[bankupper,line width=0.9pt] (-2.45,0.14)--(0,0.14);
 \draw[banklower,line width=0.9pt] (-1.55,-0.14)--(0,-0.14);
 \draw[ink,densely dashed] (-2.0,0) ellipse (1.05 and 0.8);
 \fill[ink] (-2.45,0.14) circle (1.7pt);
 \fill[ink] (-1.55,-0.14) circle (1.7pt);
 \node[ink,above] at (-2.45,0.16) {$\sigma_L$};
 \node[ink,below] at (-1.55,-0.16) {$\sigma_R$};
 \node[ink] at (-2.0,-1.2) {$\sigma_L\,\partial^m\bar\partial^m\sigma_R$};
 \node[ink] at (-1.3,-2.4) {(b)};
\end{scope}
\end{tikzpicture}
\caption{The two expansions of the two-point function of the twist fields $\sigma_L$ and $\sigma_R$ at the endpoints of the interval $[-a,b]$. (a) In the BOPE channel, we expand the 2-pt function in terms of defect operators $\widehat O_i$ on the interface. (b) After folding $\mathrm{CFT}_R$ onto $x<0$, the twist fields are at the distances $a$ and $b$ from the boundary. They are close to each other when the interval is nearly symmetric, and their product is expanded in bulk operators.}
\label{fig:twist-channels}
\end{figure}

In the replica method, the entanglement entropy is given by the 2-pt function of the replica twist fields, which can be expanded in two ways as shown in Fig.~\ref{fig:twist-channels}. There are two limits to calculate this 2-pt function: When one endpoint is much closer to the interface than the other, we expand the twist fields in defect operators on the interface. This is the BOPE, and $c_{\mathrm{eff}}$ is given by the dimension of the leading operator in it. When the interval is nearly symmetric, the two twist fields are close to each other in the folded picture, and we use their OPE in which the coefficient of the first subleading operator is fixed by $c_{LR}$ through \eqref{eq:curvature-target-expansion}. Since the two expansions represent the same function, $c_{\mathrm{eff}}-c_{LR}$ can be expressed by the data of either of them. The results are \eqref{eq:ope-difference} in the OPE channel and \eqref{eq:bope-difference} in the BOPE channel.

\paragraph{Twist fields.}
In the replica method \cite{Calabrese_2004}, $\Tr\rho^n$ for the interval $[-a,b]$ is the partition function of $n$ copies of the theory, which are connected cyclically along the interval. It is equal to the 2-pt function of the twist fields $\sigma_L$ at $x=-a$ and $\sigma_R$ at $x=b$ introduced in the main text. They are scalar primary fields of the $n$ copies of $\mathrm{CFT}_L$ and of $\mathrm{CFT}_R$, with the dimensions
\begin{equation}
 \Delta_\alpha=\frac{c_\alpha}{12}\Big(n-\frac1n\Big),\qquad\alpha=L,R.
 \label{eq:twist-dimension}
\end{equation}
The conformal transformations which map the interface to itself form the group $SL(2,\mathbb R)$, and they fix the 2-pt function up to a function of the cross-ratio $\chi=(a+b)^2/(4ab)=\cosh^2u$. We write it as
\begin{equation}
 \Tr\rho^n=\epsilon^{\Delta_L+\Delta_R}\,
 \big\langle\sigma_L(-a)\,\sigma_R(b)\big\rangle
 =\Big(\frac{\epsilon}{2a}\Big)^{\Delta_L}
  \Big(\frac{\epsilon}{2b}\Big)^{\Delta_R}e^{(1-n)F_n(u)},
 \label{eq:twist-two-point}
\end{equation}
so that the R\'enyi entropy $S_n=\frac{1}{1-n}\log\Tr\rho^n$ reduces to \eqref{eq:entropy-u} as $n\to1$, with $F(u)=\lim_{n\to1}F_n(u)$ given by the terms of the first order in $n-1$ in the expansions below (assuming that the limit $n\to1$ can be taken term by term).

\paragraph{BOPE channel.}
A twist field at the position $x$ can be expanded in terms of operators $\widehat O_i$ on the interface of the replicated theory \cite{CardyLewellen1991,wang2026twistoperatorbopeentanglement},
\begin{equation}
 \sigma_\alpha(x)=\sum_i B^\alpha_i\,|2x|^{\widehat\Delta_i-\Delta_\alpha}\,\widehat O_i+\cdots,
 \label{eq:twist-bope}
\end{equation}
where $\widehat\Delta_i$ is the dimension of $\widehat O_i$, $B^\alpha_i$ are the BOPE coefficients, and the dots stand for the derivatives of $\widehat O_i$ along the interface. The one with the lowest dimension is the boundary twist operator $\widehat\sigma$ of \cite{wang2026twistoperatorbopeentanglement}, and we denote its dimension by $\widehat\Delta_*$.

Let us first consider the BOPE channel as in Fig.~\ref{fig:twist-channels}(a). The states on the circle correspond to the operators $\widehat O_i$ and their derivatives. Summing over the derivatives with the $SL(2,\mathbb R)$ conformal block \cite{McAvityOsborn1995,LiendoRastellivanRees2013}, we obtain
\begin{equation}
 e^{(1-n)F_n(u)}=\sum_i B^L_iB^R_i\;k_{\widehat\Delta_i}(1/\chi),\qquad
 k_h(\eta)=\eta^h\,{}_2F_1(h,h;2h;\eta),
 \label{eq:bope-channel}
\end{equation}
where the operators $\widehat O_i$ are normalized to one. This is an expansion in $1/\chi=4ab/(a+b)^2$, and it converges for any $a\neq b$.

When $|u|$ is large, the sum is dominated by the boundary twist operator, $e^{(1-n)F_n(u)}\simeq B^L_*B^R_*\,(4e^{-2|u|})^{\widehat\Delta_*}$. Comparing it with $F(u)=\frac13c_{\mathrm{eff}}|u|+o(|u|)$, we see that $\widehat\Delta_*$ vanishes at $n=1$ and that the effective central charge is given by \cite{wang2026twistoperatorbopeentanglement}
\begin{equation}
 c_{\mathrm{eff}}=6\,\partial_n\widehat\Delta_*\big|_{n=1}.
 \label{eq:ceff-twist}
\end{equation}
Indeed, at $n=1$ the twist fields are the identity operator, and we have $B^L_*B^R_*=1$ and $B^L_iB^R_i=0$ for $i\neq*$. Expanding \eqref{eq:bope-channel} to the first order in $n-1$ with $\partial_hk_h(\eta)|_{h=0}=\log\eta-\frac12\log(1-\eta)$, we obtain
\begin{equation}
 F(u)=F_*+\frac{c_{\mathrm{eff}}}{6}\Big[X+\frac12\log\big(1-e^{-X}\big)\Big]
 +\sum_{i\neq*}\beta_i\,k_{\widehat\Delta_i}\big(e^{-X}\big),
 \label{eq:F-bope}
\end{equation}
in terms of the logarithmic cross-ratio $X=\log\cosh^2u$ of the main text, with
\begin{equation}
 F_*=-\partial_n\big(B^L_*B^R_*\big)\big|_{n=1},\qquad
 \beta_i=-\partial_n\big(B^L_iB^R_i\big)\big|_{n=1},
 \label{eq:bope-data}
\end{equation}
and with $\widehat\Delta_i$ in the last term evaluated at $n=1$. Notice that the saturation condition of the bound directly relates $F_*$ to $\log g$.

\paragraph{OPE channel.}
Let us next consider intervals close to the symmetric one. After folding, $\sigma_L$ and $\sigma_R$ are at the distances $a$ and $b$ from the boundary, as in Fig.~\ref{fig:twist-channels}(b), and they approach each other as $u\to0$. In the presence of the boundary only scalar operators have expectation values, and there is one such operator for each $m=0,1,2,\ldots$, of the form $\sigma_L\,\partial^m\bar\partial^m\sigma_R+\cdots$ with the dimension $\Delta_L+\Delta_R+2m$. Therefore the R\'enyi partition function reads \cite{McAvityOsborn1995,LiendoRastellivanRees2013}
\begin{equation}
 e^{(1-n)F_n(u)}=\sum_{m=0}^\infty\lambda_m\,(\chi-1)^m\,
 {}_2F_1\big(\Delta_L+m,\Delta_R+m;\Delta_L+\Delta_R+2m;1-\chi\big),
 \label{eq:ope-channel}
\end{equation}
where $\lambda_m$ is the product of the OPE coefficient and the coefficient of the 1-pt function of the $m$-th operator. The above expansion in $\chi-1=(a-b)^2/(4ab)$ vanishes for the symmetric interval. The $m=0$ term gives the interface entropy, $F(0)=-\partial_n\lambda_0|_{n=1}=\log g$.

At $n=1$ we have $\lambda_0=1$ and $\lambda_m=0$ for $m\geq1$. To the first order in $n-1$, the hypergeometric function in the $m=0$ term is $1-\frac{\Delta_L\Delta_R}{\Delta_L+\Delta_R}X$, while those in the $m\geq1$ terms can be evaluated at $n=1$, where $(\chi-1)^m{}_2F_1(m,m;2m;1-\chi)=k_m(1-e^{-X})$ with the same function $k_h$ as in \eqref{eq:bope-channel}. Since $k_1(1-e^{-X})=X$, we find for $\Phi(X)=F(u)-F(0)$,
\begin{equation}
 \Phi(X)=\bigg[\frac{c_Lc_R}{6(c_L+c_R)}-\partial_n\lambda_1\big|_{n=1}\bigg]X
 +\sum_{m=2}^\infty\phi_m\,k_m\big(1-e^{-X}\big),\qquad
 \phi_m=-\partial_n\lambda_m\big|_{n=1}.
 \label{eq:Phi-ope}
\end{equation}
Near the symmetric interval $k_m(1-e^{-X})=u^{2m}+\cdots$, so only the first term contributes to $F''(0)$. The relation \eqref{eq:curvature-target-expansion} derived in the main text is therefore equivalent to
\begin{equation}
 c_{LR}=\frac{c_Lc_R}{c_L+c_R}-6\,\partial_n\lambda_1\big|_{n=1},
 \label{eq:clr-ope}
\end{equation}
namely the transmission coefficient is determined by the first subleading operator, $m=1$, in the OPE of the twist fields. As a check, for a factorized interface, $e^{(1-n)F_n(u)}$ does not depend on $u$ and the $m=1$ term cancels the $u$-dependence of the $m=0$ term, so that $c_{LR}=0$. For the trivial interface with $c_L=c_R=c$, we have $e^{(1-n)F_n(u)}=\chi^{-\Delta_L}$, which gives $\partial_n\lambda_1|_{n=1}=-c/12$ and $c_{LR}=c$. In both cases $\phi_m=0$ for $m\geq2$.

\paragraph{The difference $c_{\mathrm{eff}}-c_{LR}$.}
We can now write out the exact expression of $c_{\mathrm{eff}}-c_{LR}$ in both expansions. With \eqref{eq:clr-ope}, the OPE channel gives
\begin{equation}
 \Phi(X)=\frac{c_{LR}}{6}X+\sum_{m=2}^\infty\phi_m\,k_m\big(1-e^{-X}\big).
 \label{eq:Phi-ope-clr}
\end{equation}
For large $X$, each term in the sum grows linearly, $k_m(1-e^{-X})\simeq\frac{\Gamma(2m)}{\Gamma(m)^2}X$, while the BOPE channel \eqref{eq:F-bope} gives $\Phi(X)=\frac16c_{\mathrm{eff}}X+O(1)$. Therefore
\begin{equation}
 c_{\mathrm{eff}}-c_{LR}=6\lim_{X\to\infty}\frac1X\sum_{m=2}^\infty\phi_m\,k_m\big(1-e^{-X}\big),
 \label{eq:ope-difference}
\end{equation}
where the sum should be performed before taking the limit.

By the convexity of $\Phi$ established in Eq.~\eqref{eq:convex} and the fact that
$\Phi(0)=0$, for any $X>0$ we have
\[
  \frac{c_{LR}}{6}=\Phi'_+(0)
  \le \frac{\Phi(X)}{X}
  \le \lim_{Y\to\infty}\frac{\Phi(Y)}{Y}
  =\frac{c_{\rm eff}}{6}.
\]
Substituting the expansion \eqref{eq:Phi-ope-clr} gives 
\begin{equation}
 0\le\sum_{m=2}^\infty\phi_m\,k_m\big(1-e^{-X}\big)\le\frac{c_{\mathrm{eff}}-c_{LR}}{6}\,X
 \qquad\text{for any }X\geq0.
 \label{eq:ope-bounds}
\end{equation}
Thus the operators with $m\geq2$ can only increase the entropy, and the bound $c_{LR}\le c_{\mathrm{eff}}$ is saturated if and only if all of them decouple at the first order in $n-1$, namely $\phi_m=0$ for $m\geq2$.

In terms of the BOPE data, we use the fact that each block in \eqref{eq:bope-channel} is an eigenfunction of the Casimir operator of $SL(2,\mathbb R)$ acting on one of the twist fields
\begin{equation}
 \big[\partial_X+(1-e^{-X})\,\partial_X^2\big]\,k_h\big(e^{-X}\big)=h(h-1)\,k_h\big(e^{-X}\big).
 \label{eq:casimir}
\end{equation}
Applying it to \eqref{eq:F-bope}, we obtain
\begin{equation}
 \Phi'(X)+(1-e^{-X})\,\Phi''(X)
 =\frac{c_{\mathrm{eff}}}{6}+\sum_{i\neq*}\widehat\Delta_i\big(\widehat\Delta_i-1\big)\,\beta_i\,k_{\widehat\Delta_i}\big(e^{-X}\big).
 \label{eq:casimir-flow}
\end{equation}
The left-hand side is equal to $\Phi'(0)=\frac16c_{LR}$ at the symmetric interval $X=0$, and it approaches $\Phi'(\infty)=\frac16c_{\mathrm{eff}}$ as $X\to\infty$, where the sum on the right-hand side vanishes. The operators on the interface other than the boundary twist operator therefore account for the difference between the two central charges, and at $X\to0$ we find
\begin{equation}
 c_{\mathrm{eff}}-c_{LR}=6\lim_{X\to0}\sum_{i\neq*}\widehat\Delta_i\big(1-\widehat\Delta_i\big)\,\beta_i\,k_{\widehat\Delta_i}\big(e^{-X}\big).
 \label{eq:bope-difference}
\end{equation}
This is the expression of $c_{\mathrm{eff}}-c_{LR}$ in terms of the BOPE of the twist fields, and our main result means that its right-hand side is non-negative for any conformal interface.

\end{document}